\documentclass[prl,showpacs,twocolumn]{revtex4}
\usepackage{amssymb}
\usepackage{amsmath}
\usepackage{graphicx}
\usepackage{epsfig}

\begin{document}

\title{Probing Quantum Phase Transition in Macroscopic Qubit Array \\
via circuit QED architecture}
\author{Y.D. Wang, Fei Xue, C.P. Sun}
\email{suncp@itp.edu.cn}
\homepage{http://www.itp.ac.cn/~suncp}
\affiliation{Institute of Theoretical Physics, Chinese Academy of Sciences, Beijing,
100080, China}
\date{\today }

\begin{abstract}
We demonstrate a universal physical mechanism to probe the
macroscopic quantum phase transition based on circuit QED
architecture. We found that, with certain parameters, the Josephson
junction qubit array behaves as an antiferromagnetic Ising model in
transverse field and the coupled transmission line resonator serves
as a bosonic quantum probe. Our investigation shows that, at the
critical point, the drastic broadening of the spectrum of the probe
indicates the quantum phase transition.
\end{abstract}

\pacs{74.81.Fa, 42.50.Pq, 75.10.Pq, 73.43.Nq}

\maketitle

\textit{Introduction-- } Non-analyticality in ground state energy of
a quantum many body system at critical point is referred to as
quantum phase transition (QPT) \cite{sachdev}, which is essentially
caused by quantum fluctuations even at zero temperature. Some recent
investigations have discovered that \cite{zurek2,quan,paz} the
critical behavior of a system with QPT can enhance quantum
decoherence of its coupled external system. Here, the enhanced
decoherence is displayed by a sensitive decay of the Loschmidt echo
(LE) \cite{Jalabert} at critical point and possesses some
universality in the ordered domain \cite{paz}. These discoveries
enlighten us to propose a scheme to probe the intrinsic QPT
phenomena of a system by detecting the exotic spectral structure of
its coupled system.

On the other hand, as a macroscopic QPT phenomenon, the superfluid-Mott
insulator transition has been demonstrated in a macroscopic quantum system
-- the atomic Bose-Einstein condensate in an optical lattice \cite{greiner}.
It is natural to extend the research on QPT to some other macroscopic
quantum systems, such as the superconducting Josephson junction (JJ) array
system. For the generic JJ array, much effort has been devoted to the
superfluid-Mott insulator transition, see Ref.\cite{zurek2} and references
therein. In "qubit" regime, this system has been studied for some other
purposes \cite{mooij,bruder, falci}, e.g., quantum state transfer.

In this paper, we investigate the macroscopic QPT of an Ising chain
in transverse field (ITF) implemented with JJ qubit array for the
first time. We present and study a physical mechanism to probe its
QPT with a coupled on-chip superconducting transmission line
resonator (TLR) \cite{yale}. We find that, when the QPT occurs in
the JJ qubit array, the spectrum of the TLR is significantly changed
from discrete-peak structure into almost white noise spectrum. This
drastic broadening of the spectrum serves as a witness of QPT. We
also discuss the universality of QPT exhibited in the spectrum
structure.

\begin{figure}[bp]
\begin{center}
\includegraphics[scale=0.9]{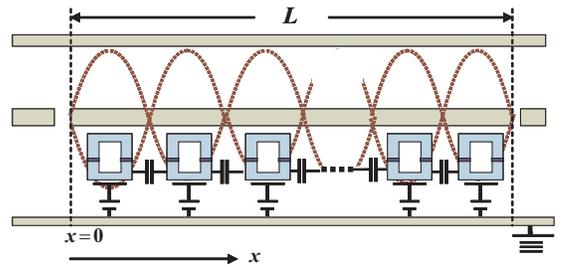}
\end{center}
\caption{(Color on line) The schematics of our setup. A capacitively coupled
Josephson junction qubit array is placed in a 1D TLR. Each qubit is coupled
with the quantized magnetic field of the TLR. }
\label{fig:circuit}
\end{figure}
\textit{QPT model based on JJ qubit array and 1D TLR -- }
We consider a quantum network including $N$ Cooper pair
boxes (CPBs) (see Fig.(\ref{fig:circuit})). Each CPB is a
dcSQUID formed by a superconducting island connected to two
Josephson junctions. The effective Josephson tunnelling
energy can be tuned by the magnetic flux $\Phi _{x}$
threading the dcSQUID. With proper bias voltage, the CPB
behaves as a qubit \cite{nec99} and then JJ qubit array
becomes an engineered ``spin" chain with $N$ $1/2$ -spins.
When the coupling
capacitance $C_{m}$ between two CPBs is much smaller than the total one $%
C_{\Sigma }$ (e.g, in ref. \cite{nec2q}, $C_{m}/C_{\Sigma }\approx 0.05$),
the terms $\sim $ $o\left( C_{m}/C_{\Sigma }\right) $ in Hamiltonian can be
neglected and we only consider the nearest neighbor interaction in this
``spin" chain. Then the JJ qubit array can be described by a 1D ITF model
with the effective Hamiltonian
\begin{equation}
H_{0}=\hat{h}\left( \lambda \right) \equiv B\sum_{\alpha=1
}^N (\lambda \sigma _{x}^{(\alpha )}+\sigma _{z}^{(\alpha
)}\sigma _{z}^{(\alpha +1)}), \label{h0}
\end{equation}
where $\lambda =B_{x}/B$ and $B=e^{2}C_{m}/C_{\Sigma }^{2}$
characterizes the Coulomb interaction between nearest
neighbors. The Josephson energy of CPB $B_{x}=E_{J}\cos
\left( \Phi _{x}/\Phi _{0}\right) /2$ with $E_{J}$ the
Josephson energy of single junction and $\Phi _{0}=h/2e$
the flux quantum. For simplicity, all qubits are assumed to
be identical and biased at the degenerate point. The
quasi-spin operators $\sigma _{z}=|0\rangle \left\langle
0\right\vert -|1\rangle \left\langle 1\right\vert $,
$\sigma _{x}=-|0\rangle \left\langle 1\right\vert
-|1\rangle \left\langle 0\right\vert $ are defined in terms
of the charge eigenstates $\left\vert 0\right\rangle $ and
$\left\vert 1\right\rangle $. $\left\vert 0\right\rangle $
and$\left\vert 1\right\rangle $ denote $0$ and $1$ excess
Cooper pair on the island respectively. A most recent
experiment has demonstrated the possibility to implement a
four-JJ-qubit Ising array \cite{4q}.

In our setup, as a quantum probe, a 1D TLR of length $L$ is placed in
parallel with this JJ qubit array (see Fig.(\ref{fig:circuit})) away from a
distance $d$. Each CPB situates at the antinodes $x=(2n+1)L/2N$ ($n=0$, $%
\cdots $, $N-1 $) of the magnetic field induced by the curent $J$ in the TLR
\cite{yale}. Since $J$ vanishes at the end of the TLR, the London equation
provides the boundary condition for the electromagnetic field of this
on-chip resonator. Thus, the electric field vanishes at those antinodes and
the qubits are only coupled with the magnetic component. The magnetic flux
threading each dcSQUID is $\phi _{x}=\eta \left( a+a^{\dag }\right) $ with $%
\eta =(S/d)\left( \hbar l\omega /L\right) ^{1/2}$ where $l$ is the
inductance per unit length and $S$ is the enclosed area of the dcSQUID.
Here, we have assumed only a single mode of magnetic field with frequency $%
\omega $ is coupled with JJ qubit array \cite{singlemode} and $a$ ($a^{\dag }
$) is its annhilation (creation) operator. Usually, $\eta $ is small enough
for the harmonic approximation \cite{zurek2,ydwang} $\cos \phi
_{x}\thickapprox 1-\phi _{x}^{2}$ and the Hamiltonian $H=H_{0}+H_{F}$ takes
a spin-boson form
\begin{equation}
H_{F}=\hbar \omega a^{\dag }a-g\sum_{\alpha }\left( a^{\dag }a+aa^{\dag
}\right) \sigma _{x}^{(\alpha )},
\end{equation}
with the coupling coefficient $g=\eta E_{J}$. Here, we have already invoked
the rotation wave approximation to neglect the high frequency terms
proportional to $a^{\dag }a^{\dag }$ and $a^{2}$ under the condition $\omega
>>B_{x}$, $B$. This approximation condition can be satisfied with accessable
parameters in current experiments. For example, if we take $C_{\Sigma }\sim
600$ aF and $C_{\Sigma }\sim 30$ aF, $L\sim 1$ cm, $S\sim 10$ $\mu $m$^{2}$,
$d\sim 1$ $\mu $m and $N=500$, then $B=1.6$ GHz, $E_{J}=13$ GHz, $\omega
\sim 120$ GHz, $\eta \sim 0.01$ \cite{Wang2005}.

\textit{Pseudo-spin representation for paired excitation spectrum - } By
introducing the Jordan-Wigner transformation $\sigma _{z}^{(\alpha
)}=\prod_{\beta <\alpha }\left( 2c_{\beta }^{\dag }c_{\beta }-1\right)
\left( c_{\alpha }+c_{\alpha }^{\dag }\right) $ and $\sigma _{x}^{(\alpha
)}=1-2c_{\alpha }^{\dag }c_{\alpha }$, $H_{0}$ can be diagonalized as $%
H_{0}=\sum_{k}\varepsilon _{k}\gamma _{k}^{\dag }\gamma _{k}$ by the
fermionic quasi-particle operator \cite{sachdev,pfeuty,lieb}
\begin{equation}
\gamma _{k}=\sum_{\alpha =1}^{N}\frac{e^{-ik\alpha }}{\sqrt{N}}(c_{\alpha
}\cos \frac{\theta _{k}}{2}-ic_{\alpha }^{\dag }\sin \frac{\theta _{k}}{2})
\end{equation}
with dispersion relation $\varepsilon _{k}(\lambda )=2B\sqrt{1+\lambda
^{2}-2\lambda \cos k}$ for $\tan \theta _{k}(\lambda )=\sin k/\left( \lambda
-\cos k\right)$. The ground state $\left\vert G\right\rangle $ of $H_{0}$
describes the state without any quasi-particle excitation.

With respect to the Fock state $\left\vert n\right\rangle $ of TLR, the
Hamiltonian of the whole system can be decomposed as $H=\sum_{n}H_{n}\left%
\vert n\right\rangle \left\langle n\right\vert $ where $H_{n}=\hat{h}\left(
\lambda _{n}\right) $ are defined by Eq.(\ref{h0}) with $\lambda
_{n}=\lambda -\left( 2n+1\right) g/B$ and a constant term $\hbar n\omega $
has been omitted. For further convenience, we introduce a set of pseudo-spin
operators \cite{Anderson}
\begin{eqnarray}
s_{zk} &=&\gamma _{k}^{\dag }\gamma _{k}+\gamma _{-k}^{\dag }\gamma _{-k}-1,
\notag \\
s_{xk} &=&i\left( \gamma _{-k}\gamma _{k}+\gamma _{-k}^{\dag }\gamma
_{k}^{\dag }\right) ,  \notag \\
s_{yk} &=&\gamma _{-k}^{\dag }\gamma _{k}^{\dag }-\gamma _{-k}\gamma _{k}.
\end{eqnarray}
They describe the pairing of quasi-particle excitations by
$\gamma _{k}$. With these pseudo-spin operators, each
branch Hamiltonian $H_{n}$ can be rewritten as
$H_{n}=\sum_{k>0}H_{n}^{(k)}$, where
\begin{equation}
H_{n}^{(k)}=\varepsilon _{nk}(s_{zk}\cos 2\alpha _{nk}+s_{xk}\sin 2\alpha
_{nk})
\end{equation}%
with $2\alpha _{nk}=\theta _{nk}-\theta _{k}$, $\varepsilon
_{nk}=\varepsilon _{k}(\lambda _{n})$ and $\theta _{nk}=\theta _{k}(\lambda
_{n})$.

\textit{Detection of QPT --} We expect to detect the critical behavior of
the JJ qubit array by the coherence property of the TLR. To demonstrate the
quantum coherence of a single mode electromagnetic field, a natural option
is the correlation spectrum function $S\left( \omega \right) =\int
dte^{-i\omega t}S\left( t\right) $, which is the Fourier transformation of
the $1$st order correlation function of the single mode field
\begin{equation}
S(t)=\left\langle a^{\dag }\left( t\right) a\left( 0\right)
\right\rangle =\sum_{n}n\left\vert c_{n}\right\vert
^{2}D_{n,n-1}\left( t\right) e^{-\Gamma \left\vert t\right\vert }.
\label{st}
\end{equation}
Here, the average $\left\langle \cdots \right\rangle $ is taken over
an initial state $\left\vert \Psi \left( 0\right) \right\rangle
=\left\vert \psi _{0}\right\rangle \otimes \left\vert G\right\rangle
$ and $\left\vert \psi _{0}\right\rangle =\sum_{n}c_{n}\left\vert
n\right\rangle $ is an arbitrary pure state of the TLR (our
discussion here is also valid if $\left\vert \psi _{0}\right\rangle$
is an arbitrary mixed state). The decoherence factor
$D_{n,n-1}\left( t\right) =\left\langle G\left\vert \exp
(iH_{n}t)\exp (-iH_{n-1}t)\right\vert G\right\rangle$ evaluates the
overlap of the wave functions under two different Hamiltonians
$H_{n}$ and $H_{n-1}$. We also phenomenologically introduce the
decaying factor $\exp (-\Gamma \left\vert t\right\vert )$ in the
quasi-mode treatment of dissipation \cite{singlemode}. For strong
coupling limit, $g\gg \Gamma $ and $\Gamma $ is about $6.3$ MHz for
the first excitation mode \cite{yale}.

By carrying out the evaluation of time evolution, we obtain explicitly the
spectrum function
\begin{equation}
S\left( \omega \right) =\sum_{n}n\left\vert
c_{n}\right\vert ^{2}D_{n,n-1}\left( \omega \right),
\label{sw}
\end{equation}%
where
\begin{equation}
D_{n,n-1}\left( \omega \right) =\sum_{\left\{ \left(
a_{k},b_{k}\right) \right\} }\frac{2\Gamma F_{{( a_{k},b_{k})} }^{(
n) }}{\Gamma ^{2}+(\omega -\Omega _{\left\{ \left(
a_{k},b_{k}\right) \right\} }^{\left( n\right) })^{2}}  \label{dw1}
\end{equation}%
is the sum of many Lorentzian distributions with the same half width
at half maximum (HWHM) $\Gamma$, but different central frequencies
$\Omega _{\left\{ \left( a_{k},b_{k}\right) \right\}
}^{(n)}=\sum_{k}\left( a_{k}\varepsilon _{nk}+b_{k}\varepsilon
_{n-1,k}\right) $. The sum in eq.( \ref{dw1}) is taken over all the
possible configurations of combinations $ \left\{ \left(
a_{k},b_{k}\right) |a_{k},b_{k}=\pm \right\} $, e.g., one possible
combination is $\left\{ \left( +,-\right) _{1},\left( +,+\right)
_{2},\cdots \left( -,+\right) _{N/2}\right\} $. Here, $F_{\left\{
\left( a_{k},b_{k}\right) \right\} }^{\left( n\right)
}=\prod_{k}c_{a_{k}b_{k},k}^{\left( n,n-1\right) }$ is defined by
\begin{eqnarray} \nonumber
c_{++,k}^{\left( n,n-1\right) } &=&-\sin \alpha _{nk}\cos
\alpha _{n-1k}\sin \left( \alpha _{n-1k}-\alpha
_{nk}\right) , \\ \nonumber
c_{+-,k}^{\left( n,n-1\right) }
&=&\sin \alpha _{nk}\sin \alpha _{m-1k}\cos \left( \alpha
_{n-1k}-\alpha _{nk}\right) , \\ \nonumber
c_{-+,k}^{\left(
n,n-1\right) } &=&\cos \alpha _{nk}\cos \alpha _{n-1k}\cos
\left( \alpha _{n-1k}-\alpha _{nk}\right) , \\
c_{--,k}^{\left( n,n-1\right) } &=&\cos \alpha _{nk}\sin \alpha _{n-1k}\sin
\left( \alpha _{n-1k}-\alpha _{nk}\right) .
\end{eqnarray}%
Without considering the decay of quasimodes, those Lorentzian line shapes
reduce to delta functions.

\begin{figure}[bp]
\centering
\includegraphics[bb=53 637 265 761,scale=1, clip]{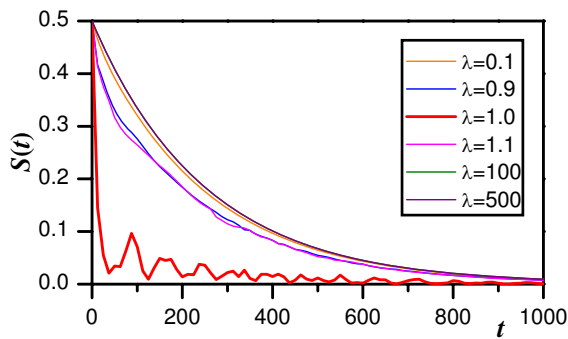}
\caption{(Color on line) The 1st order correlation function
$S(t)$ for $|\psi_0\rangle=(|0\rangle +|1\rangle
)/\sqrt{2}$ is plotted with different $\lambda $. Here
$N=1000$ and the time $t$ is in the unit of $1/B$. }
\label{fig:st}
\end{figure}

The time evolution of the $1$st order correlation function $S\left(
t\right) $ with $N=1000$ is shown in Fig.(\ref{fig:st}) for
different $\lambda $ with $\left\vert \psi _{0}\right\rangle =\left(
\left\vert 0\right\rangle +\left\vert 1\right\rangle \right)
/\sqrt{2}$. It can be seen that the decay rates for different
$\lambda $ are almost the same except $\lambda =1$. This decay is
induced by the dissipation of the quasimodes, which has the same
influence for different $\lambda $. However, near the critical
point, i.e., $\lambda \approx 1$, the decay is drastically enhanced.
This means that there exists an extra strong decay mechanism related
to QPT.
\begin{figure}[tp]
\centering
\includegraphics[bb=53 548 294 760,scale=0.95, clip]{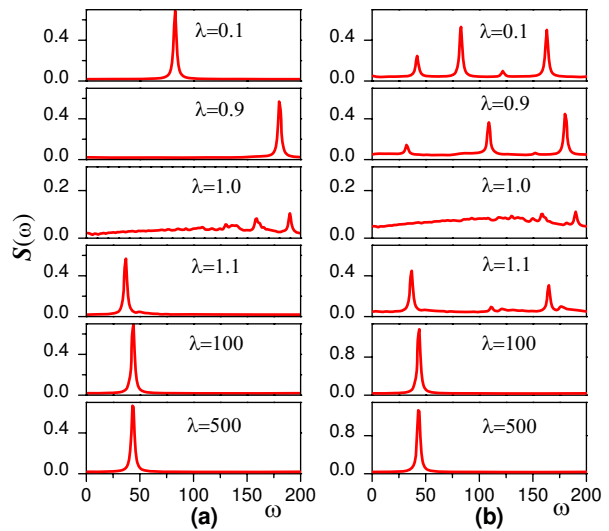}
\caption{(Color on line) The spectrum $S(\omega )$ is shown with
different $\lambda $. The left panel is plotted for $|\psi
_{0}\rangle =(|0\rangle +|1\rangle )/\sqrt{2}$ and the right panel
is for $|\psi _{0}\rangle =|\alpha \rangle $. Here $N=1000$.}
\label{fig:sw1}
\end{figure}

\begin{figure}[tp]
\centering
\includegraphics[bb=55 614 248 760,scale=1, clip]{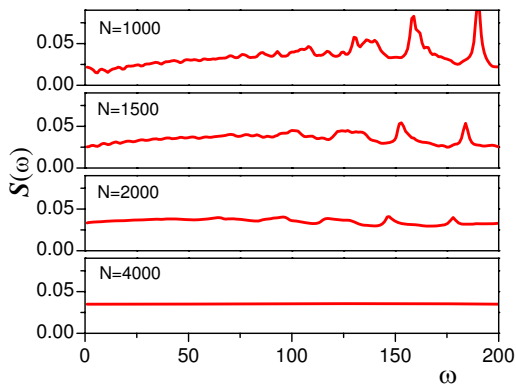}
\caption{(Color on line) The spectrum $S(\omega )$ at the critical
point is plotted for different $N$ with $|\psi_0\rangle =(|0\rangle
+|1\rangle )/\sqrt{2}$. } \label{fig:sw2}
\end{figure}

To illustrate the effect of QPT more clearly, we resort to the
behavior of the spectrum function $S(\omega )$. The numerical result
by FFT is shown in Fig.(\ref{fig:sw1}) and (\ref{fig:sw2}). In
Fig.(\ref{fig:sw1}), the left panel is plotted for $\left\vert \psi
_{0}\right\rangle =\left( \left\vert 0\right\rangle +\left\vert
1\right\rangle \right) /\sqrt{2}$ while the right panel for
$\left\vert \psi _{0}\right\rangle =\left\vert \alpha \right\rangle
$ with $\alpha =1$. It can be seen that, for both of the two initial
states, generally there are only one or several Lorentzian peaks
centered at discrete frequencies in $S\left( \omega \right)$ while
near the phase transition point the spectrum of TLR gets broad and
chaotic. As $N$ increases, this broadened distribution at the
critical point becomes more and more smooth and tends to be a white
noise spectrum at large $N$ limit (see Fig.(\ref{fig:sw2})). Thus,
the QPT of the JJ qubit array is featured by the intensive
broadening in the TLR output spectrum. Hence, from the correlation
spectrum of the quantum probe, we can infer the occurrence of QPT.

To investigate the underlying physical mechanism for the behavior described
above, we rewrite $S\left( \omega \right) $ as
\begin{equation}
S\left( \omega \right) =\sum_{n,ii^{\prime }}p_{i,i^{\prime }}^{(n)}\langle
E_{i}^{\left( n\right) }|E_{i^{\prime }}^{\left( n-1\right) }\rangle
L(\omega ,\Lambda _{ii^{\prime }}^{(n)},\Gamma ).
\end{equation}%
Here $|E_{i}^{\left( n\right) }\rangle $ is an eigenvector of $H_{n}$ with
eigenvalue $E_{i}^{\left( n\right) }$, $p_{i,i^{\prime }}^{(n)}=n\left\vert
c_{n}\right\vert ^{2}\langle G|E_{i}^{\left( n\right) }\rangle \langle
E_{i^{\prime }}^{\left( n-1\right) }|G\rangle $ and
\begin{equation}
L(\omega ,\Lambda _{ii^{\prime }}^{(n)},\Gamma )=\frac{2\Gamma }{\Gamma
^{2}+\left( \omega -\Lambda _{ii^{\prime }}^{(n)}\right) ^{2}}
\end{equation}%
is a Lorentzian function with the HWHM $\Gamma $ and central frequency $%
\Lambda _{ii^{\prime }}^{(n)}=E_{i}^{\left( n\right) }-E_{i^{\prime
}}^{\left( n-1\right) }$. In some sense, $S(\omega )$ measures how
many different eigenvectors of $H_{n-1}$ are needed to express one
eigenvector of $H_{n}$. The more are necessary, the wider the
support of $S\left( \omega \right) $. This observation provides the
intrinsic reason for the widening of the spectrum. Since $g$ is
assumed to be small perturbation one would generally expects that
the difference between $H_{n}$ and $H_{n-1}$ is almost negligible
and their eigenvectors are very close to each other, that is
$\langle E_{i}^{\left( n\right) }|E_{i^{\prime }}^{\left( n-1\right)
}\rangle \approx \delta _{i,i^{\prime }}$ and
\begin{equation}
S\left( \omega \right) \approx L\left( \omega ,0,\Gamma \right)
\sum_{n}n\left\vert c_{n}\right\vert ^{2}  \label{dw2}
\end{equation}%
Then the support of $S\left( \omega \right) $ is very narrow and the
corresponding Fourier transformation $S\left( t\right) $ decays very slow.

However, the above analysis is invalid at the critical point. Near
the critical point, the property of the QPT system, such as ground
state and long range order is significantly influenced by a small
perturbation in either of the two competing terms: the Ising
interaction and the transverse field. The seemingly very small
difference between the two Hamiltonians $H_{n}$ and $H_{n-1}$
actually has drastic impact on the evolutions driven by the two
Hamiltonians. This implies that more eigenvectors of $H_{n-1}$ are
needed to reproduce one eigenvector of $H_{n}$. Therefore, more
Lorentzian shapes have to be included and the support of $S(\omega
)$ becomes much broader. This in turn accelerates the decay of
$S(t)$ and acts as the extra strong decay mechanism related to QPT
as we have noticed in Fig.(\ref {fig:st}).

The mechanism described in the paper can be extended to the case
with the two-level atom as probe. The universality similar to Ref
\cite{paz} is also revealed in Fig.(\ref {fig:sw1}). Here, when
$\lambda$ is not large, the location of peaks in $ S(\omega )$
depends on both $\lambda $ and $|\psi _{0}\rangle $. However, for
very large $\lambda $, there is only one Lorentzian peak in the
spectrum and the location of this peak is independent of $\lambda $
and $\left\vert \psi _{0}\right\rangle $. In Fig.(\ref {fig:st}), we
can also see that the decay envelope for $\lambda=500$ overlaps with
$\lambda=100$. This is because the approximation in eq.(\ref{dw2})
is rigorously hold only if the JJ qubit array is far away from the
critical point. In this case, the spectrum exhibits universal
features.

This probe mechanism requires $g\gg \Gamma$, which ensures the
decoherence related to QPT is far more prominent than that caused by
surrounding environment. But $g$ also should be much smaller than
the energy scale of the free qubit array. Otherwise, the QPT nature
of the ITF model would be significantly changed.

\textit{Conclusion --} In this paper, with superconducting circuit
QED structure, we demonstrate a detection scheme for the macroscopic
QPT phenomenon. By examining the coherent output of the coupled TLR,
the quantum criticality of the JJ qubit array can be probed. The
developing experiments \cite{nec2q,yale,4q} make our scheme to be
potentially feasible in the near future. Concerning experimental
implementation, we would like to point out only the case of
$B_{z}=0$ is discussed here to obtain an analytical result. But due
to the unavoidable charge fluctuation in our system, it is hard to
set the bias charge to $1/2$ precisely. Therefore, more realistic
consideration reminds us to concern the case of $B_{z}\neq 0$, which
is modeled with a transverse field Ising model also with a
longitudinal field. When the longitudinal field is weak enough, this
generalized model near critical point can be revealed with a
perturbation theory.

This work is funded by NSFC with grant Nos. 90203018, 10474104, 60433050,
and NFRPC with Nos. 2001CB309310 and 2005CB724508.


\begin{thebibliography}{99}
\bibitem{sachdev} S. Sachdev, \textit{Quantum Phase Transition, }(Cambridge
University Press, Cambridge, 1999).

\bibitem{zurek2} J. Dziarmaga, A. Smerzi, W.H. Zurek and A. R. Bishop, Phys.
Rev. Lett. \textbf{88}, 167001 (2002).

\bibitem{quan} H.T. Quan, Z. Song, X. F. Liu, P. Zanardi and C.P. Sun, Phys.
Rev. Lett. \textbf{96}, 140604 (2006).

\bibitem{paz} F.M. Cucchietti, S. Fernandez-Vidal, J.P. Paz,
quant-ph/0604136.

\bibitem{Jalabert} R.A. Jalabert and H.M. Pastawski, Phys. Rev. Lett.
\textbf{86}, 2490 (2001).

\bibitem{greiner} M. Greiner, O. Mandel, T. Esslinger, \textit{et al.}, Nature \textbf{415}, 39 (2002).

\bibitem{mooij} L.S. Levitov, T.P. Orlando, J.B. Majer and J.E. Mooij,
cond-mat/0108266.

\bibitem{bruder} A. Romito, R. Fazio and C. Bruder, Phys. Rev. B \textbf{71}
, 100501(R) (2005); A. Lyakhov and C. Bruder, New J. Phys.
\textbf{7}, 181 (2005).

\bibitem{falci} M. Paternstro, G.M. Palma, M.S. Kim and G. Falci, Phys. Rev.
A \textbf{71}, 042311 (2005).

\bibitem{yale} A. Wallraff, \textit{et al.}, Nature \textbf{431}, 162
(2004); A. Blais, \textit{et al.}, Phys. Rev. A \textbf{69}, 062320
(2004).

\bibitem{nec99} Y. Nakamura, Yu. A. Pashkin and J. S. Tsai, Nature
\textbf{398}, 786 (1999).

\bibitem{nec2q} Yu. A. Pashkin, \textit{et al.} Nature (London), \textbf{421}
, 823 (2003); T. Yamamoto, Y.A. Pashkin, O. Astafiev, \textit{et
al.} ,Nature (London), \textbf{425}, 941 (2003).

\bibitem{4q} M. Grajcar \textit{et al.}, Phys. Rev. Lett. \textbf{96}, 047006 (2006).

\bibitem{singlemode} For very large $N$, the energy spectrum of the cavity
mode is quasi-continuous. In principle it is hard to single out one
mode especially when the two systems are not exactly resonant. But
if we takethe dissipation for the cavity into account, there are
only some discrete Fox-Li quasimodes surrounded by many addtional
modes. The additional modes induce the decay of the Fox-Li quasimode
with decay rate $\Gamma $. Therefore, the single mode approximation
is still hold. Even though, it is worth to point out that our
proposal here is also valid for multimode field.

\bibitem{ydwang} Y. D. Wang, P. Zhang, D. L. Zhou, and C. P. Sun, Phys. Rev.
B \textbf{70}, 224515 (2004)

\bibitem{Wang2005} Y.D. Wang, Z.D. Wang and C.P. Sun, Phys. Rev. B
\textbf{72}, 172507 (2005).

\bibitem{pfeuty} P. Pfeuty, Ann. Phys.(N.Y.) \textbf{57}, 79 (1970).

\bibitem{lieb} E. Lieb, T. Schultz and D. Mattis, Ann. Phys. (N. Y.) 16, 407
(1961).

\bibitem{Anderson} P. W. Anderson, Phys. Rev. \textbf{112,} 1900(1958).
\end{thebibliography}
\end{document}